Research Article

# Analytical Study of the Holographic Superconductor from Higher Derivative Theory

Chunyan Wang,[1] Dan Zhang,[1] Guoyang Fu [ID],[2] and Jian-Pin Wu [ID],[2,3]

[1]*Department of Physics, School of Mathematics and Physics, Bohai University, Jinzhou 121013, China*
[2]*Center for Gravitation and Cosmology, College of Physical Science and Technology, Yangzhou University, Yangzhou 225009, China*
[3]*School of Aeronautics and Astronautics, Shanghai Jiao Tong University, Shanghai 200240, China*

Correspondence should be addressed to Jian-Pin Wu; jianpinwu@mail.bnu.edu.cn

In this paper, we analytically study the holographic superconductor models with the high derivative (HD) coupling terms. Using the Sturm-Liouville (S-L) eigenvalue method, we perturbatively calculate the critical temperature. The analytical results are in good agreement with the numerical results. It confirms that the perturbative method in terms of the HD coupling parameters is available. Along the same line, we analytically calculate the value of the condensation near the critical temperature. We find that the phase transition is second order with mean field behavior, which is independent of the HD coupling parameters. Then, in the low-temperature limit, we also calculate the conductivity, which is qualitatively consistent with the numerical one. We find that the superconducting energy gap is proportional to the value of the condensation. But we note that since the condensation changes with the HD coupling parameters, as the function of the HD coupling parameters, the superconducting energy gap follows the same change trend as that of the condensation.

## 1. Introduction

The mechanism of the high-temperature superconductor is one of the long-standing important and fundamental issues in strongly correlated condensed matter physics. AdS/CFT (Anti-de Sitter/Conformal Field theory) correspondence [1–4] provides a powerful tool and novel mechanism to attack this problem. Great progresses have been made, and the first holographic superconductor model has been constructed in [5]. This model exhibits appealing characteristics, one of which is the superconducting energy gap $\omega_g/T_c \approx 8$. This value roughly approximates that measured in high-temperature superconductor materials [6]. It is in contrast to the one of $\omega_g/T_c \approx 3.5$ from weakly coupled BCS theory.

Lots of works on the holographic superconductor have been fully explored (see [7–9] and references therein). An interesting holographic superconductor model is first constructed in one class of 4-derivative theory [10], which involves the coupling between the Maxwell field and a Weyl tensor, and after that, lots of extended studies on the holographic superconductor in 4-derivative theory framework are explored in [11–21]. Compared with the usual holographic superconductor, the superconducting energy gap $\omega_g/T_c$ runs from 6, approaching the one of weakly coupled BCS theory, to 10, which is beyond that of the usual holographic superconductor and shed a light on the study of the high-temperature superconducting energy gap [10, 11, 20] (The running of superconducting energy gap is also observed in the Gauss-Bonnet (GB) gravity [22, 23] and the quasitopological gravity [24, 25], in which $\omega_g/T_c$ is always greater than 8.). Furthermore, the holographic superconductor and its properties, including the superconducting energy gap and Homes' law, have also been studied in 6-derivative theory, where the Maxwell fields couple more Weyl tensor [26]. There are wider superconducting energy gap proximately ranging from 5.5 to 16. And in certain range of parameters of the holographic superconductor models in both 4- and 6-derivative theory frameworks, the experimental results of Homes' law can be satisfied.

Most of the studies are implemented numerically. To back up the numerical results and especially explore the mechanism behind these phenomena, we can resort to the analytical methods. The analytical matching method is developed in [22, 23, 27–29] to calculate the critical temperature and critical magnetic field. But the validity depends on the choice of the matching point. A particular case is that the analytical method is invalid for the GB holographic superconductor with zero scalar mass since the curvature correction term does not contribute to the analytic approximation [23]. A new analytic procedure by using the second-order Sturm-Liouville (S-L) method is developed to work out the critical temperature in [30], in which the analytic result is in good agreement with the numerical one. And then, this procedure is widely applied to other holographic models and proved to be powerful (see [31–49] and references therein). In addition, the optical conductivity can be also worked out analytically [30]. By this way, the superconducting energy gap can be estimated analytically, which is in good agreement with the numerical result.

In this paper, we will analytically study the holographic superconductor model from higher derivative (HD) theory [10, 26] by using the S-L variational method. The plan of this work is organized as follows. In Section 2, we present a brief introduction of the holographic superconductor models from HD theory. By using the S-L method, we analytically work out the critical temperature of the superconducting phase transition in Section 3. In Section 4, the condensation near the critical temperature is also analytically obtained. Then, we analytically derive the low frequency conductivity in the low-temperature limit in Section 5. In particular, the superconducting energy gap is approximately worked out. Finally, the conclusion and discussion are presented in Section 6.

## 2. The Holographic Superconductor from HD Theory

In this section, we present a brief review on the holographic superconductor from HD theory. For more details, please refer to [26]. We shall work in the probe limit. So, we shall start with the background geometry of the 4-dimensional SS-AdS black brane:

$$ds^2 = -f(r)dt^2 + \frac{1}{f(r)}dr^2 + \frac{r^2}{L^2}\left(dx^2 + dy^2\right),$$
$$f(r) = \frac{r^2}{L^2}\left(1 - \frac{r_+^3}{r^3}\right), \quad (1)$$

where $r_+$ is the horizon of the black brane while the conformal AdS boundary locates at $r \longrightarrow \infty$. The Hawking temperature of this black brane is

$$T = \frac{3r_+}{4\pi L^2}. \quad (2)$$

On top of the above fixed background, we consider the actions including the charged complex scalar field $\Psi$ and the Maxwell field strength $F_{\mu\nu}$ coupling with the Weyl tensor as follows:

$$S_A = \int d^4x \sqrt{-g}\left(-\frac{L^2}{8g_F^2} F_{\mu\nu} X^{\mu\nu\rho\sigma} F_{\rho\sigma}\right), \quad (3)$$

$$S_\Psi = \int d^4x \sqrt{-g}\left(-|D_\mu \psi|^2 - m^2 |\psi|^2\right). \quad (4)$$

In the action $S_A$, $F_{\mu\nu} = \nabla_\mu A_\nu - \nabla_\nu A_\mu$ with $A_\mu$ being the $U(1)$ gauge field. The tensor $X$ comprises an infinite family of HD terms [50] (The HD coupling terms are introduced by the Weyl tensor, which vanishes at the conformal AdS boundary. Such HD coupling terms can give lots of interesting results. For instance, they break the electromagnetic (EM) self-duality in 4 dimensional spacetimes, which corresponds to the particle-vortex duality in the boundary theory [51–56]. In addition, the similar coupling terms can be also introduced to construct the holographic quantum critical phase (QCP) [57–59]. When the backreaction is included, we can construct the metal insulator phase transition [55, 60] and explore the effect on the chaos, the holographic entanglement, the thermodynamics, and the holographic thermalization from the Weyl tensor [60–64]. We can also construct the holographic superconductors with the HD couplings of an $U(1)$ field to a scalar field as in [65–67]. This type of holographic superconductor model exhibits a rich phase structure, and in particular, the transition from the normal to the superconducting phase can be tuned to be of first order or of second order depending of the coupling parameters.):

$$X_{\mu\nu}{}^{\rho\sigma} = I_{\mu\nu}{}^{\rho\sigma} - 8\gamma_{1,1} L^2 C_{\mu\nu}{}^{\rho\sigma} - 4L^4 \gamma_{2,1} C^2 I_{\mu\nu}{}^{\rho\sigma}$$
$$- 8L^4 \gamma_{2,2} C_{\mu\nu}{}^{\alpha\beta} C_{\alpha\beta}{}^{\rho\sigma} - 4L^6 \gamma_{3,1} C^3 I_{\mu\nu}{}^{\rho\sigma}$$
$$- 8L^6 \gamma_{3,2} C^2 C_{\mu\nu}{}^{\rho\sigma} - 8L^6 \gamma_{3,3} C_{\mu\nu}{}^{\alpha_1\beta_1} C_{\alpha_1\beta_1}{}^{\alpha_2\beta_2} C_{\alpha_2\beta_2}{}^{\rho\sigma} + \cdots, \quad (5)$$

where $I_{\mu\nu}{}^{\rho\sigma} = 2\delta_\mu{}^{[\rho}\delta_\nu{}^{\sigma]}$ is an identity matrix and $C^n = C_{\mu\nu}{}^{\alpha_1\beta_1} C_{\alpha_1\beta_1}{}^{\alpha_2\beta_2} \cdots C_{\alpha_{n-1}\beta_{n-1}}{}^{\mu\nu}$. The factor of $L$ in Equations (3) and (5) is introduced so that the coupling parameters $g_F$ and $\gamma_{i,j}$ are dimensionless. But for later convenience, we shall set $L = 1$ and $g_F = 1$ below. When $X_{\mu\nu}{}^{\rho\sigma} = I_{\mu\nu}{}^{\rho\sigma}$, the action $S_A$ is just the standard Maxwell theory. For convenience, we write $\gamma_{1,1} = \gamma$ and $\gamma_{2,i} = \gamma_i$ ($i = 1, 2$). In this paper, we shall truncate the $X$ tensor up to the square of the Weyl tensor, which constructs 4- or 6-derivative theory. Since the effect of both 6-derivative terms, i.e., $\gamma_1$ and $\gamma_2$ terms, is similar [26, 50], it is enough to study $\gamma$ and $\gamma_1$ terms in this paper. Note that in the SS-AdS black brane background, the parameters $\gamma$ and $\gamma_1$ are constrained in the regions $-1/12 \leq \gamma \leq 1/12$ [51, 68] and $\gamma_1 \leq 1/48$ [50], respectively, when other parameters are turned off. Here, we also study the superconductivity in these parameter spaces. $\Psi$ in the action $S_\Psi$ is the charged complex scalar field, which has mass $m$ and charge $q$ of the gauge field $A$. It is convenient to write $\Psi$ as $\Psi = \psi e^{i\theta}$, where $\psi$ and $\theta$ are the real scalar field and the Stückelberg field, respectively. The covariant derivative $D_\mu$ is defined as $D_\mu = \partial_\mu - iqA_\mu$. Choosing the gauge $\theta = 0$, we obtain the EOMs of the gauge field and scalar field as

$$\nabla_\nu \left( X^{\mu\nu\rho\sigma} F_{\rho\sigma} \right) - 4q^2 A^\mu \psi^2 = 0, \quad (6)$$

$$\left[ \nabla^2 - \left( m^2 + q^2 A^2 \right) \right] \psi = 0. \quad (7)$$

Further, we assume the following form for both the gauge field and the scalar field as

$$\begin{aligned} A &= (\phi(r), 0, 0, 0), \\ \psi &= \psi(r), \end{aligned} \quad (8)$$

which are only the function of the radial coordinate $r$. Under this ansatz, the EOMs (6) can be explicitly written as

$$\psi''(r) + \frac{4r^3 - r_+^3}{r^4 - rr_+^3} \psi'(r) - \frac{m^2 \left( r^4 - rr_+^3 \right) - r^2 \phi^2(r)}{(r^3 - r_+^3)^2} \psi(r) = 0,$$

$$\phi''(r) + \frac{2\left( r^6 + 4r^3 r_+^3 \gamma + 96 r_+^6 \gamma_1 \right)}{r^7 - 8r^4 r_+^3 \gamma - 48 r r_+^6 \gamma_1} \phi'(r)$$

$$- \frac{2q^2 r^7 \psi^2(r)}{(r^3 - r_+^3)(r^6 - 8r^3 r_+^3 \gamma - 48 r_+^6 \gamma_1)} \phi(r) = 0, \quad (9)$$

where the prime denotes the derivative with respect to $r$. It is more convenient to work in the coordinate $u = r_+/r$, in which $u = 1$ and $u = 0$ are the horizon and the AdS boundary, respectively. Then, the above EOMs become

$$\psi''(u) - \frac{2 + u^3}{u - u^4} \psi'(u) + \frac{m^2 r_+^2 (u^3 - 1) + q^2 u^2 \phi^2(u)}{r_+^2 u^2 (u^3 - 1)^2} \psi(u) = 0, \quad (10)$$

$$\phi''(u) - \frac{24 u^2 (\gamma + 12 u^3 \gamma_1)}{1 - 8u^3 \gamma - 48 u^6 \gamma_1} \phi'(u) - \frac{2q^2 \psi^2(u)}{u^2(1 - u^3)(1 - 8u^3 \gamma - 48 u^6 \gamma_1)} \phi(u) = 0, \quad (11)$$

where the prime now represents the derivative with respect to $u$. We shall fix the mass of the scalar field to $m^2 = -2$ in this paper. So, the asymptotical behaviors for both $\psi$ and $\varphi$ at the boundary $u \longrightarrow 0$ are

$$\phi = \mu - \frac{\rho}{r_+} u, \quad (12)$$

$$\psi = u \psi_1 + u^2 \psi_2, \quad (13)$$

where $\mu$ and $\rho$ are interpreted as the chemical potential and charge density of the dual boundary field theory, respectively. According to the AdS/CFT correspondence, either $\psi_1$ or $\psi_2$ will act as a condensation operator while the other will be identified as a source. Here, we treat $\psi_1$ as the source and $\psi_2$ as the expectation value, which is denoted as $\psi_2 = \langle \mathcal{O}_+ \rangle$.

## 3. The Critical Temperature

In this section, following the SL method proposed in [30], we analytically work out the critical temperature $T_c$ for the superconducting phase transition. Note that through this paper, we set $q = 1$.

To this end, we consider the system approaching the phase transition point, for which $T \longrightarrow T_c$ such that we can approximately set $\psi = 0$. Then, the Maxwell EOM (11) reduces to

$$\phi''(u) - \frac{24 u^2 \left( \gamma + 12 u^3 \gamma_1 \right)}{1 - 8u^3 \gamma - 48 u^6 \gamma_1} = 0. \quad (14)$$

Since the introduction of the HD coupling term, it is hard to directly obtain the analytic solution of $\phi(u)$ as that in [30]. Alternatively, by treating the coupling parameters $\gamma$ and $\gamma_1$ as the small quantities, we can solve perturbatively the above equation order by order. (Since for the allowed region of the coupling parameters, the denominator $1 - 8u^3\gamma - 48u^6\gamma_1$ in Equation (14) is larger than zero, it is safe to solve perturbatively this equation in terms of the order of the coupling parameters.) Up to the first order of $\gamma$ and $\gamma_1$, we obtain $\phi(u)$ as

$$\phi(u) = -(1-u)\mathscr{C} - 2\gamma(1-u^4)\mathscr{C} - \frac{48}{7}\gamma_1 \mathscr{C}(1-u^7), \quad (15)$$

where $\mathscr{C}$ is a constant of integration. When $\gamma = \gamma_1 = 0$, $\phi(u) = C(u - 1)$, which reduces to that without the Weyl coupling term as [30]. The integration constant $\mathscr{C}$ can be determined by the UV asymptotic behavior of $\phi(u)$ (Equation (12)), which gives

$$\mathscr{C} = -\frac{\rho}{r_+}. \quad (16)$$

It is convenient to introduce $\lambda = \rho/r_{+c}^2$, for which $r_{+c}$ is the radius of the horizon at $T = T_c$. Then, the solution for $\phi(u)$ (15) can be rewritten as

$$\phi(u) = \lambda r_{+c} (1-u) \left[ 1 + 2\gamma \xi(u) + \frac{48\gamma_1}{7} \xi_1(u) \right], \quad (17)$$

where $\xi(u) = 1 + u + u^2 + u^3$ and $\xi_1(u) = 1 + u + u^2 + u^3 + u^4 + u^5 + u^6$.

After the solution of $\phi(u)$ is at hand, we turn to solve Equation (10) for the scalar field $\psi(u)$. Near the critical temperature ($T \longrightarrow T_c$), Equation (10) becomes

$$\psi''(u) + \frac{u^3 + 2}{u(u^3 - 1)} \psi'(u) + \frac{1}{u^3 - 1}$$
$$\cdot \left[ -\frac{2}{u^2} + \frac{\lambda^2(u-1)(7 + 14\gamma\zeta(u) + 48\gamma_1\zeta_1(u))^2}{49(u^2 + u + 1)} \right] \psi(u) = 0, \tag{18}$$

where we have used the solution of $\phi(u)$, i.e., Equation (17). In order to match the behavior at the conformal AdS boundary, we introduce the form for $\psi(u)$ as

$$\psi(u) = \frac{\langle \mathcal{O}_+ \rangle}{\sqrt{2} r_+^2} u^2 F(u). \tag{19}$$

$F(u)$ is introduced in the above equation as a trial function, for which we shall take the formula as

$$F(u) = 1 - \alpha u^2, \tag{20}$$

where $\alpha$ is the parameter to be determined. It satisfies the boundary conditions $F(0) = 1$ and $F'(0) = 0$. Inserting Equation (19) into Equation (18), one obtains a second-order S-L self-adjoint differential equation for $F(u)$ as

$$(u^2 - u^5) F''(u) + (2u - 5u^4) F'(u) - 4u^3 F(u)$$
$$- \frac{\lambda^2 u^2 (7 + 14\gamma + 48\gamma_1 - 7u - 14\gamma u^4 - 48\gamma_1 u^7)^2}{49(u^3 - 1)} F(u) = 0. \tag{21}$$

The S-L theory indicates that the eigenvalue $\lambda^2$ minimizes the expression [37]:

$$\lambda^2 = \frac{\int_0^1 \left\{ T(u) [F'(u)]^2 + P(u)[F(u)]^2 \right\} du}{\int_0^1 Q(u)[F(u)]^2 du}, \tag{22}$$

where

$$T(u) = u^2 - u^5,$$
$$P(u) = 4u^3,$$
$$Q(u) = -\frac{u^2 (7 + 14\gamma + 48\gamma_1 - 7u - 14\gamma u^4 - 48\gamma_1 u^7)^2}{49(u^3 - 1)}. \tag{23}$$

Then, the critical temperature $\widehat{T}_c \equiv T_c/\sqrt{\rho}$ can be given as

$$\widehat{T}_c = \frac{3}{4\pi\sqrt{\lambda}}. \tag{24}$$

The strategy for calculating the critical temperature is to minimize the expression (22) of the eigenvalue $\lambda^2$ with respect to the coefficient $\alpha$, and then, substituting the value of $\lambda$ into Equation (24), we can obtain the value of $\widehat{T}_c$.

When the HD coupling terms vanish, i.e., $\gamma = \gamma_1 = 0$, the eigenvalue $\lambda^2$ can be analytically worked out as

$$\lambda^2 \big|_{\gamma=0, \gamma_1=0}$$
$$= \frac{48\alpha^2 - 80\alpha + 60}{10\sqrt{3}\pi(\alpha^2 - 1) + 30(3 - \ln(3)) - 3\alpha^2(7 + 10\ln(3)) + \alpha(130 - 60\ln(9))}. \tag{25}$$

Its minimum can be achieved at $\alpha \approx 0.6016$. The corresponding minimum value of $\lambda^2$ is $\lambda^2 \approx 17.309$. Therefore, the critical temperature $\widehat{T}_c$ reads as $\widehat{T}_c = 0.11704$. This analytical result is in very good agreement with the numerical result: $\widehat{T}_c = 0.118$ [5].

But when $\gamma \neq 0$ or $\gamma_1 \neq 0$, it is hard to work out the analytical expression of $\lambda^2$. Therefore, we resort to the perturbative calculation. For the 4-derivative term, we obtain the value of $\lambda^2$ up to the order of $\gamma$ as $\mathcal{O}(\gamma^6)$, while for the 6-derivative term, we obtain its value up to the order of $\gamma_1$ as $\mathcal{O}(\gamma_1^8)$. Finally, we present the results for 4- and 6-derivative theories in Tables 1 and 2, respectively. For comparison, we also give the corresponding numerical results in Tables 1 and 2. From the two tables, we see that the analytical results we worked out here are in very good agreement with the numerical ones. It confirms that the perturbative method to calculate the value of $\lambda^2$ is available.

Table 1: The analytical and numerical results for the critical temperature of the holographic superconductors from the 4-derivative theory.

| $\gamma$ | $-1/12$ | $-1/24$ | 0 | $1/24$ | $1/12$ |
|---|---|---|---|---|---|
| $\widehat{T}_c\big|_{\text{Analytical}}$ | 0.103 | 0.109 | 0.117 | 0.129 | 0.149 |
| $\widehat{T}_c\big|_{\text{Numerical}}$ | 0.105 | 0.111 | 0.118 | 0.130 | 0.150 |

Table 2: The analytical and numerical results for the critical temperature of the holographic superconductors from the 6-derivative theory.

| $\gamma_1$ | $1/48$ | 0 | $-1/48$ |
|---|---|---|---|
| $\widehat{T}_c\big|_{\text{Analytical}}$ | 0.170 | 0.117 | 0.106 |
| $\widehat{T}_c\big|_{\text{Numerical}}$ | 0.183 | 0.118 | 0.107 |

## 4. Condensation

In this section, we aim to investigate the condensation operator near the phase transition region. Away from

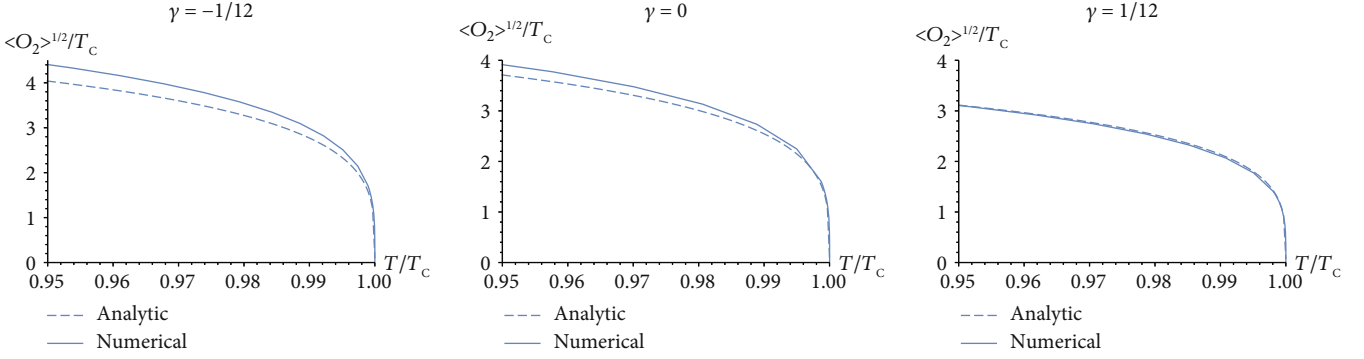

FIGURE 1: The condensation from 4-derivative theory for different $\gamma$ obtained analytically and numerically, respectively.

(but close to) the critical temperature $T_c$, field Equation (11) for $\phi$ can be rewritten as

$$\phi''(u) + \frac{24u^2\gamma}{(8u^3\gamma - 1)}\phi'(u) - \frac{\langle\mathcal{O}_+\rangle^2 u^2 F^2(u)}{r_+^4(u^3 - 1)(8u^3\gamma - 1)}\phi(u) = 0. \quad (26)$$

Near the phase transition region, the condensation $\langle\mathcal{O}_+\rangle^2/r_+^4$ is very small. Thus, we can expand $\phi(u)$ perturbatively in terms of the condensation $\langle\mathcal{O}_+\rangle^2/r_+^4$ as

$$\frac{\phi(u)}{r_+} = \lambda(1-u)\left[1 + 2\gamma\xi(u) + \frac{48}{7}\gamma_1\xi_1(u)\right] + \frac{\langle\mathcal{O}_+\rangle^2}{r_+^4}\chi(u)+,\cdots, \quad (27)$$

where $\chi(u)$ satisfies the boundary condition $\chi(1) = \chi'(1) = 0$. Substituting Equation (27) into Equation (26), we obtain the differential equation for $\chi(u)$ up to the first order of $\langle\mathcal{O}_+\rangle^2/r_+^4$:

$$\left[(1 - 8u^3\gamma - 48u^6\gamma_1)\chi'(u)\right]' = \lambda\frac{u^2(7 + 14\gamma\xi(u) + 48\gamma_1\xi_1(u))}{7(u^2 + u + 1)}F(u)^2. \quad (28)$$

Integrating both sides of the above equation in the interval [0,1] with the boundary condition, we obtain

$$\chi'(0) = -\lambda\mathcal{A}, \quad (29)$$

where

$$\mathcal{A} = \int_0^1 \frac{u^2(7 + 14\gamma\xi(u) + 48\gamma_1\xi_1(u))}{7(u^2 + u + 1)}F(u)^2 du. \quad (30)$$

Near the boundary $u = 0$, combining Equations (12) and (27), one has

$$\frac{\mu}{r_+} - \frac{\rho}{r_+^2}u = \lambda(1-u)\left[1 + 2\gamma\xi(u) + \frac{48}{7}\gamma_1\xi_1(u)\right] + \frac{\langle\mathcal{O}_+\rangle^2}{r_+^4}\left[\chi(0) + u\chi'(0)+\cdots\right]. \quad (31)$$

Comparing the coefficient of $u$ from both sides of the equation, we obtain

$$\frac{\rho}{r_+^2} = \lambda - \frac{\langle\mathcal{O}_+\rangle^2}{r_+^4}\chi'(0) = \lambda\left(1 + \frac{\langle\mathcal{O}_+\rangle^2}{r_+^4}\mathcal{A}\right). \quad (32)$$

Therefore, the condensate near the critical temperature is

$$\langle\mathcal{O}_+\rangle = \kappa T_c^2\sqrt{1 - \frac{T}{T_c}}, \quad (33)$$

where

$$\kappa = \left(\frac{4\pi}{3}\right)^2 \frac{2}{\sqrt{\mathcal{A}}}. \quad (34)$$

It indicates that the phase transition is second order with mean field behavior, which is independent of the HD coupling parameters. It analytically confirms the results numerically obtained in our previous works [10, 26].

Finally, we also give the analytical and numerical condensation for samples $\gamma$ and $\gamma_1$ in Figures 1 and 2, respectively. Near the critical temperature, the analytical results match well with the numerical ones.

## 5. The Conductivity at Low-Temperature Region

In this section, we turn to study the conductivity at the low-temperature region. When temperature $T \longrightarrow 0$, the dominant contribution comes from the neighborhood of the boundary ($u = 0$). So, we make a rescaling $\psi(bu)$ and $\phi(bu)$, and then, $u \longrightarrow u/b$ with letting $b \longrightarrow \infty$. Thus, the field Equations (10) and (11) can be simplified as

$$F''(u) + \frac{2}{u}F'(u) + \frac{\phi^2(u)}{r_+^2}F(u) = 0, \tag{35}$$

$$\phi''(u) - \frac{\langle O_+ \rangle^2 u^2}{r_+^4} F^2(u)\phi(u) = 0, \tag{36}$$

where we have used Equation (19) and restored the original coordinate $u$. We shall solve the above equations subject to the following boundary conditions:

$$3F'(1) + 4F(1) = 0, \quad F(0) = 1, \; F'(0) = 0. \tag{37}$$

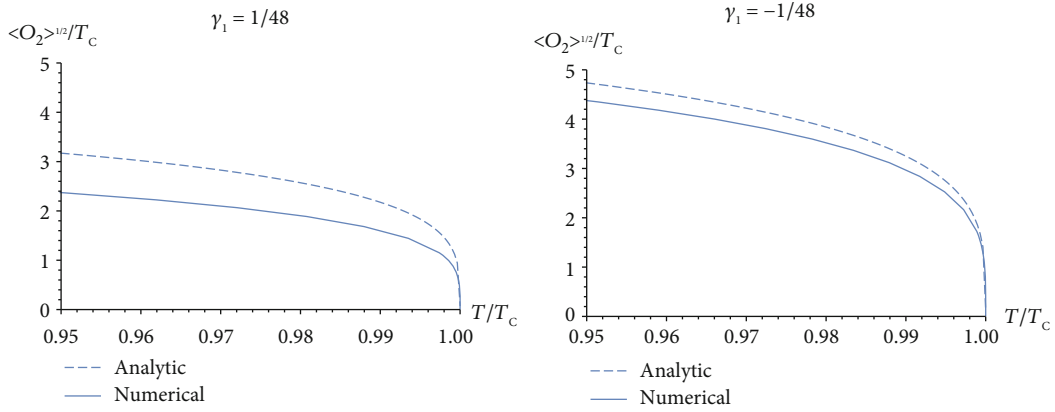

FIGURE 2: The condensation from 6-derivative theory for different $\gamma_1$ obtained analytically and numerically, respectively.

Equation (37) is the natural boundary condition at the horizon, which can be directly deduced from Equation (21).

To proceed, we assume that in the asymptotic regime closing to the conformal boundary, $F(u)$ takes the following power law formula:

$$F(u) \approx \frac{\beta}{bu}, \tag{38}$$

where $\beta$ is the parameter to be determined. Then, the solution of field Equation (36) for $\phi(u)$ is

$$\phi(u) = \mathscr{B} r_+ \sqrt{bu} K_{1/2}(bu), \tag{39}$$

$$b^2 = \frac{\langle O_+ \rangle \beta}{r_+^2}, \tag{40}$$

where $\mathscr{B}$ is the integral constant and $K_n(z)$ is the modified Bessel function of the second kind with $n$ denoting the order of the corresponding Bessel function. Near the boundary $u = 0$, using Equation (12), the ratio $\rho/r_+^2$ can be estimated as

$$\frac{\rho}{r_+^2} = -\frac{\Gamma(-(1/2))}{2^{3/2}}\mathscr{B}b = \sqrt{\frac{\pi}{2}}\mathscr{B}b. \tag{41}$$

Substituting Equation (39) into Equation (35) and rescaling $u \longrightarrow u/b$, we have

$$F'' + \frac{2}{u}F' + \tilde{\mathscr{B}}^2 u(K_{1/2}(u))^2 F = 0, \quad \tilde{\mathscr{B}} = \frac{\tilde{\mathscr{B}}}{b}. \tag{42}$$

The above equation is the second-order S-L self-adjoint differential equation. It should be solved in the interval $(0, \infty)$ subject to the boundary condition $F(0) = 1$, $F'(0) = 0$, and $F \longrightarrow 0$ as $u \longrightarrow \infty$. The expression for estimating the minimum eigenvalue of $\tilde{\mathscr{B}}$ is provided by

$$\tilde{\mathscr{B}}^2 = \frac{\int_0^\infty \left[uF'(u)\right]^2 du}{\int_0^\infty u^3 K_{1/2}^2(u) F^2(u) du}. \tag{43}$$

In order to connect smoothly the asymptotic regime depicted by the power law formula (38) with the boundary condition $F(0) = 1$, we introduce the following trial function:

$$F_\beta(u) = \left(\frac{\beta}{u}\right) \tanh\left(\frac{u}{\beta}\right). \tag{44}$$

Then, we can minimize expression (43) to fix the value of $\beta$. It is easy to find that we have the minimum $\tilde{\mathscr{B}} \approx 1.92$ at $\beta \approx 0.8$.

Once the solution for the scalar field $\psi$ is at hand, we can turn to study the optical conductivity. To this end, we switch on the perturbations of the gauge field along $x$ direction, $A_x(r, t) = A(r)e^{-i\omega t}$. And then, the perturbative gauge field equation can be read as

$$A''(u) + \frac{3u^2\left(1 - \gamma(4 - 8u^3)\right) - 48\gamma_1\left(3u^3 - 2\right)u^3}{(u^3 - 1)(1) + 4\gamma u^3 - 48\gamma_1 u^6} A'(u)$$
$$+ \left[\frac{\omega^2}{r_+^2(u^3 - 1)^2} + \frac{2\psi(u)^2}{u^2(u^3 - 1)(1 + 4\gamma u^3 - 48\gamma_1 u^6)}\right] A(u) = 0. \tag{45}$$

To proceed, we reexpress the above equation as the following form:

$$\left[r_+(1-u^3)(1+4\gamma u^3 - 48\gamma_1 u^6)A'(u)\right]'$$
$$+ \frac{\omega^2(1+4\gamma u^3 - 48\gamma_1 u^6)}{r_+(1-u^3)}A(u) = \frac{2r+\psi^2(u)}{u^2}A(u). \quad (46)$$

We want to recast the above equation as a Schrödinger form. So, we move to the tortoise coordinate, which is defined by

$$r_* = \int \frac{dr}{f(r)} = \frac{1}{6r_+}\left[\ln\frac{(1-u)^3}{1-u^3} - 2\sqrt{3}\tan^{-1}\frac{\sqrt{3}u}{u+2}\right]. \quad (47)$$

The integration constant has been calculated from the boundary condition that $r_* = 0$ at $u = 0$. Then, Equation (46) becomes

$$\frac{d^2\Phi}{dr_*^2} + \left[\omega^2 - V_{\text{eff}}\right]\Phi = 0, \quad (48)$$

where $\Phi(u) = \sqrt{1+4\gamma u^3 - 48\gamma_1 u^6}A(u)$, and the effective potential $V_{\text{eff}}$ is

$$V_{\text{eff}} = \frac{2f\psi^2}{-48\gamma_1 u^6 + 4\gamma u^3 + 1} + \frac{6u^3\left(2\gamma^2(7u^3-1)u^3 - \gamma(5u^3-2)(144\gamma_1 u^6 - 1) + 24\gamma_1 u^3(240\gamma_1 u^9 - 96\gamma_1 u^6 - 8u^3 + 5)\right)}{(1+4\gamma u^3 - 48\gamma_1 u^6)^2}f. \quad (49)$$

The wave Equation (48) is to be solved subject to ingoing boundary condition at the horizon, i.e., the $\omega$-dependent part of the equation for $V_{\text{eff}} = 0$. The solution reads

$$\Phi(u) \sim e^{-i\omega r_*} \sim (1-u)^{-i\omega/3r_+}. \quad (50)$$

So, at low frequency, to account for the boundary condition at the horizon, we define

$$A = \frac{(1-u)^{-i\omega/3r_+}}{\sqrt{1+4\gamma u^3 - 48\gamma_1 u^6}}\mathcal{G}(u), \quad (51)$$

where $\mathcal{G}(u)$ is regular at the horizon ($u = 1$). Then, Equation (45) becomes

$$\left[\frac{18u(24\gamma_1 u^3(30\gamma u^6 + (8-12\gamma)u^3 + 48\gamma_1(2-5u^3)u^6 - 5) + \gamma(-14\gamma u^6 + (2\gamma-5)u^3 + 2))}{(-48\gamma_1 u^6 + 4\gamma u^3 + 1)^2}\right]\mathcal{G}$$
$$+ \left[\frac{6\psi(u)^2}{u^2(1+4\gamma u^3 - 48\gamma_1 u^6)} - \frac{i(2u+1)\omega}{r_+} - \frac{(u+2)(u^2+u+4)\omega^2}{3r_+^2(u^2+u+1)}\right]\mathcal{G} - 3(1-u^3)\mathcal{G}'' + \left[9u^2 - 2(1+u+u^2)\frac{i\omega}{r_+}\right]\mathcal{G}' = 0. \quad (52)$$

We can expand the wave function $\mathcal{G}(u)$ in a Taylor series at the horizon, which gives the boundary condition as

$$(3r_+ - 2i\omega)\mathcal{G}'(1) + \left[\frac{2r_+(-9\gamma + 216\gamma_1 + \psi(1)^2)}{4\gamma - 48\gamma_1 + 1} - \frac{2\omega^2}{3r_+} - i\omega\right]\mathcal{G}(1) = 0. \quad (53)$$

At low temperature, it is convenient to solve Equation (52) by letting $u \longrightarrow u/b$ with $b \longrightarrow \infty$. Then, Equation (52) becomes

$$3\mathcal{G}''(u) + \frac{2i\omega}{r_+}\mathcal{G}'(u) - \left[3b^2\tanh\left(\frac{bu}{\beta}\right)^2 - \frac{8\omega^2}{3r_+^2}\right]\mathcal{G}(u) = 0. \quad (54)$$

The general solution of Equation (54) can be given in terms of the Legendre function:

$$\mathcal{G}(u) \approx \left(\frac{1-\tanh(bu/\beta)}{1+\tanh(bu/\beta)}\right)^{i\omega\beta/6br_+}$$
$$\cdot \left[c_+ P^{+\beta}_{-(1/2)+(1/2)\sqrt{1+4\beta^2}}\left(\tanh\frac{bu}{\beta}\right)\right. \quad (55)$$
$$\left. + c_- P^{-\beta}_{-(1/2)+(1/2)\sqrt{1+4\beta^2}}\left(\tanh\frac{bu}{\beta}\right)\right],$$

where $c_+$ and $c_-$ are the integration constants, for which we shall derive them in terms of the boundary condition at the horizon below.

At $u=1$, we have $\tanh(bu/\beta) \approx 1$, the Legendre functions become

$$P^{\pm\beta}_{-(1/2)+(1/2)\sqrt{1+4\beta^2}}\left(\tanh\frac{bu}{\beta}\right) \simeq \frac{2^{\pm\beta/2}}{\Gamma(1\mp\beta)}\left(1-\tanh\frac{bu}{\beta}\right)^{\mp(\beta/2)}. \quad (56)$$

Therefore, we obtain

$$\mathcal{G}(1) \approx \left[\frac{c_+}{\Gamma(1-\beta)}e^{+b} + \frac{c_-}{\Gamma(1+\beta)}e^{-b}\right]e^{-(i\omega/3r_+)},$$
$$\mathcal{G}'(1) \approx \left[\frac{c_+(b-(i\omega/3r_+))}{\Gamma(1-\beta)}e^{+b} - \frac{c_-(b+(i\omega/3r_+))}{\Gamma(1+\beta)}e^{-b}\right]e^{-(i\omega/3r_+)}. \quad (57)$$

Substituting $\mathcal{G}(1)$ and $\mathcal{G}'(1)$ in the boundary condition (53), we obtain the ratio $(c_+/c_-)$ as

$$\frac{c_+}{c_-} = -e^{-2b}\frac{\Gamma(1-\beta)}{\Gamma(1+\beta)}\mathcal{M}, \quad (58)$$

where

$$\mathcal{M} = \left[\frac{(b^4-3b)-6(3+2b)\gamma+144(3+b)\gamma_1}{(b^4+3b)-6(3-2b)\gamma+144(3-b)\gamma_1}\right. $$
$$\left. + \frac{4i\omega b(1+4\gamma-48\gamma_1)(b^4-30\gamma-3+576\gamma_1)}{r_+(b^4+3b(1+4\gamma-48\gamma_1)-18(\gamma-24\gamma_1))^2}\right]. \quad (59)$$

According the definition of conductivity and AdS/CFT correspondence, one has

$$\sigma(\omega) \approx i\frac{\sqrt{\langle\mathcal{O}_+\rangle}}{\omega}\frac{0.47-0.66(c_+/c_-)}{0.85-0.30(c_+/c_-)}. \quad (60)$$

So, in the limit of the low frequency ($\omega \longrightarrow 0$), we have

$$\Re\sigma(\omega) \sim e^{-2b} = e^{-E_g/T},$$
$$\Im\sigma(\omega) \approx 0.55\frac{\sqrt{\langle\mathcal{O}_+\rangle}}{\omega}. \quad (61)$$

$E_g$ is identified to be the superconducting energy gap, which is determined as

$$E_g \approx \frac{3\sqrt{\beta\langle\mathcal{O}_+\rangle}}{2\pi} \approx 0.43\sqrt{\langle\mathcal{O}_+\rangle}. \quad (62)$$

This result reveals that the superconducting energy gap is proportional to the value of the condensation. But we note that since the condensation changes with the HD coupling parameters, as the function of the HD coupling parameters, the superconducting energy gap follows the same change trend as that of the condensation. This analytical result is qualitatively consistent with the numerical one in [10, 26].

Since the introduction of the HD term, which complicates the EOM, we cannot analytically obtain the conductivity as the function of $\omega$ at $T \longrightarrow$ as [29]. In the future, we can seek new methods to do this thing. For example, we can study the conductivity as the function of $\omega$ at zero temperature by using the semianalytical methods as [69].

## 6. Conclusions and Discussions

In this paper, we have analytically studied the holographic superconductor models from the HD theory. To achieve this goal, we use the Sturm-Liouville (S-L) eigenvalue method, which has been widely used in holographic models. Different from the usual holographic superconductor models, we cannot derive the analytical expression for the eigenvalue $\lambda^2$ due to the introduction of the HD coupling terms. Instead, we develop the perturbative method in terms of the HD coupling parameters to calculate the eigenvalue $\lambda^2$ and so the critical temperature. We find that the analytical results are in good agreement with the numerical results, which confirms that the perturbative method is available. Along the same line, we calculate the value of the condensation near the critical temperature. We find that the phase transition is second order with mean field behavior, which is independent of the HD coupling parameters. It analytically confirms the results numerically obtained in our previous works [10, 26]. We also calculate the conductivity in the low-temperature limit, which is qualitatively consistent with the numerical one [10, 26]. We find that the superconducting energy gap is proportional to the value of the condensation. But we note that since the condensation changes with the HD coupling parameters, as the function of the HD coupling parameters, the superconducting energy gap follows the same change trend as that of the condensation.

## Data Availability

No data were used to support this study.

## Conflicts of Interest

The authors declare that they have no conflicts of interest.

## Acknowledgments

This work is supported by the Natural Science Foundation of China under Grant Nos. 11775036 and 11847313. This work is also supported by the "Top Talent Support Programme from Yangzhou University."

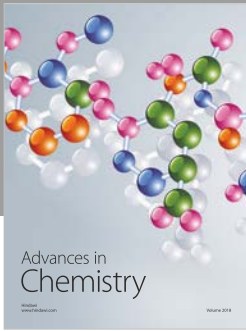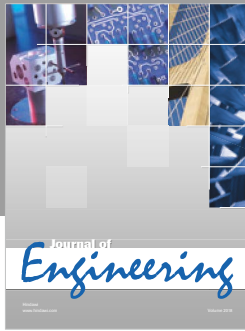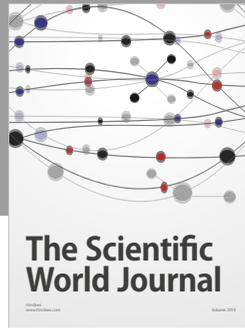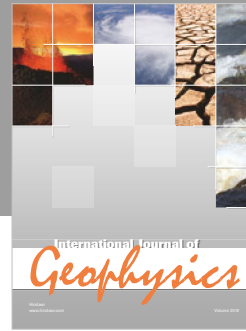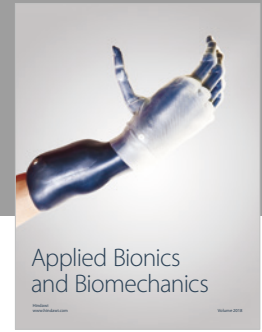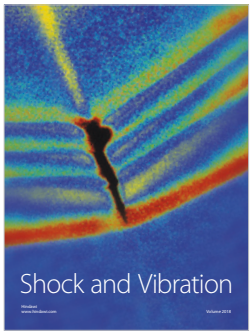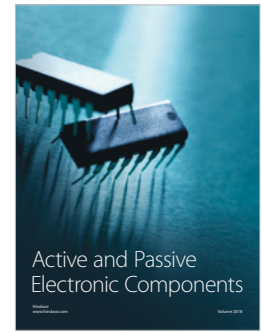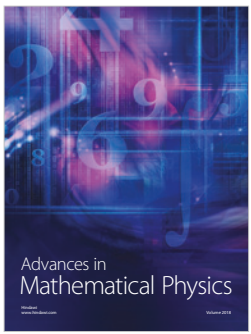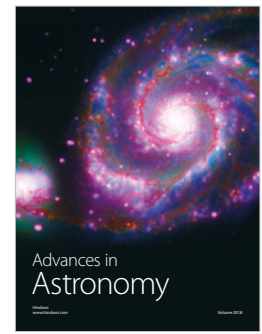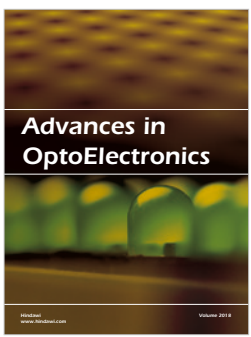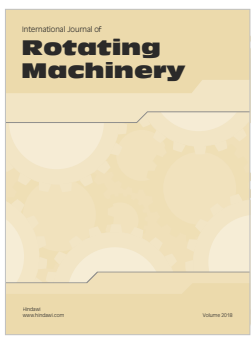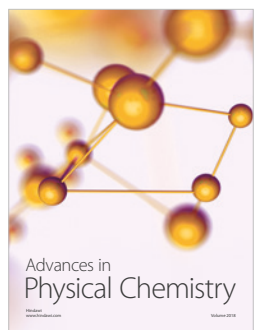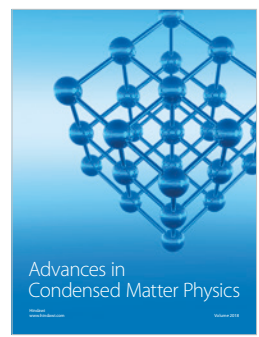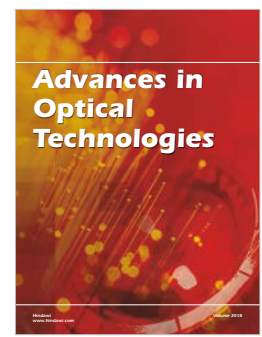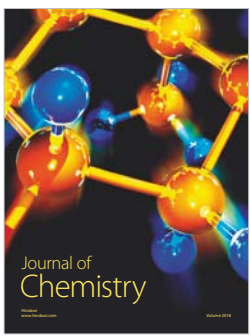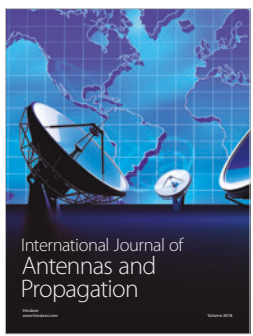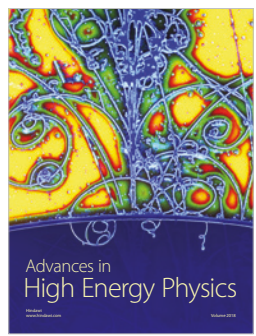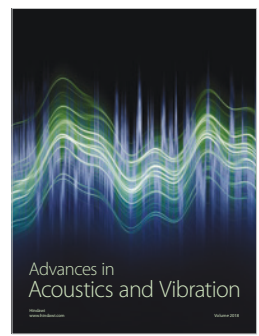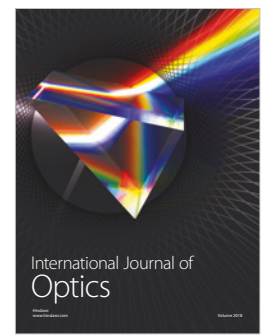

Submit your manuscripts at
www.hindawi.com